\documentclass[useAMS,usenatbib]{mn2e}

\usepackage{graphicx}
\usepackage{amsmath}
\usepackage{amssymb}
\usepackage{color}
\usepackage{gensymb}
\usepackage{subfigure}

\title[High velocity stars from GC-SMBHB interaction]{High velocity stars from the interaction of a globular cluster and a massive black hole binary}
\author[G. Fragione and R. Capuzzo-Dolcetta]{G. Fragione$^{1}$\thanks{E-mail: giacomo.fragione@uniroma1.it} and R. Capuzzo-Dolcetta$^{1}$\thanks{E-mail:
roberto.capuzzodolcetta@uniroma1.it}\\
$^{1}$Dep. of Physics, Sapienza, Univ. of Roma, P.le A. Moro 2, 00185, Roma, Italy}
\begin{document}



\maketitle

\begin{abstract}
High velocity stars are stars moving at velocities so high to require an acceleration mechanism involving binary systems or the presence of a massive central black hole. In the frame of a galaxy hosting a supermassive black hole binary (of total mass $10^8$ M$_\odot$), we investigated a mechanism for the production of high velocity stars due to the close interaction between a massive and orbitally decayed globular cluster and the super massive black hole binary. Some stars of the cluster acquire high velocities by conversion of gravitational energy into kinetic energy deriving from their interaction with the black hole binary. After the interaction, few stars reach a velocity sufficient to overcome the galactic gravitational well, while some of them are just stripped from the globular cluster and start orbiting around the galactic centre.
\end{abstract}


\label{firstpage}

\section{Introduction}

Massive Black Holes (BHs) are likely in the majority of galactic nuclei over the whole Hubble sequence and are recognized as fundamental building blocks in models of galaxy formation and evolution \citep{kor13,dub14,ses14}. According to the present cosmological model $\Lambda$CDM, structures formation involves hierarchical mergers of galaxies, which follow the hierarchical growth of their parent dark matter halos \citep{may07,ros15}. Galaxies may experience multiple mergers during their lifetime and, if more than one of these interacting galaxies host a massive BH, the formation of a Black Hole Binary (BHB) is a natural consequence of the hierarchical paradigm \citep*{beg80,vol03}. How long BHBs survive and whether they merge are key points for several questions in high-energy and extragalactic astronomy, as the detection of gravitational waves \citep{ses05,ses13,rss14}.

BHBs lifetime is governed by the loss of energy and angular momentum, by means of which two massive BHs become gravitationally bound and finally merge. BHBs are thought to undergo four dynamical stages before their coalescence \citep*{beg80,yu02}. The first one is the dynamical friction stage, during which each BH inspirals independently toward the centre of the common gravitational potential on the Chandrasekhar time-scale \citep{cha43}

\begin{equation}
t_{df}\sim\frac{4\times 10^6}{\log N}\left(\frac{\sigma_c}{200\ km\ s^{-1}}\right)\left(\frac{r_c}{100\ pc}\right)^2\left(\frac{10^8 M_{\odot}}{m_i}\right) yr\ ,
\end{equation}

where $N$, $\sigma_c$, $r_c$ are the number of stars, the one-dimensional velocity dispersion and the radius of the core, respectively, while $m_i$ ($i=1,\ 2$) is the BH mass. During the second stage, usually referred to as non-hard binary stage, BHs velocities increase, while their orbital period shortens, because more and more stars in the galactic nuclear core are scattered off through gravitational slingshots, while the dynamical friction becomes less efficient. The third stage, labelled hard binary stage, begins when the orbital separation $a$ of BHs is about their influence radius \citep{qui96}

\begin{equation}
a_h=2.8\left(\frac{m_2}{10^8 M_{\odot}}\right)\left(\frac{200\ km\ s^{-1}}{\sigma_c}\right)^2 pc\ ,
\end{equation}

where $m_2$ is the mass of the lighter BH. Hard BHBs lose energy mainly by the three-body slingshot effect with stars passing in their vicinity, which are expelled after one or more encounters. The span of this phase depends on the loss-cone refilling. If it is dominated by two-body relaxation in a spherical and symmetric system, energy will not be lost efficiently and the binary may stall at parsec scale (the so-called "final parsec problem"). However, more realistic calculations show that this stall does not sussist when, for example, asymmetric potentials or gas dynamics are taken into account \citep*{mer05,ber06,mer07,kha13}. Finally, the last stage is characterised by the energy loss due to gravitational radiation. A BHB will coalesce within the time \citep{pet64}

\begin{equation}
t_{coal}\sim\frac{5.8\ m_1^2\times 10^6}{m_2 M}\left(\frac{a}{0.01\ pc}\right)^4\left(\frac{10^8 M_{\odot}}{m_1}\right)^3,
\end{equation} 

where $m_1$ is the mass of the heavier BH and $M=m_1+m_2$ is the total mass.

On the point of view of observations, Super Massive BHBs (SMBHBs) are quite hard to detect, in particular in dry environments. However, the presence of surrounding gas enhances the possibilities of finding and resolving BHBs in galactic centres thanks to their energetic interactions \citep*{kom03,rod06,fab11,dea14}. Moreover, in these gas-rich systems, the BHB evolution is likely to be driven primarily by gas \citep{goi15}. On the other hand, in gas-poor environments, when tidal disruption events enlighten the cores of galaxies hosting quiescent BHBs, observers are able to detect and study them \citep*{che09,liu09,che11,weg11,liu13,liu14}.

Many galaxies show nucleated central regions, the so-called Nuclear Star Clusters (NSCs), which are among the densest stellar populations observed in the Universe \citep*{mat97,car97,car98,cot06,tur12}. Two different processes are thought to give birth to NSCs. The first mechanism involves radial gas inflow into the galactic centre and predicts that NSCs consist mostly of stars formed locally \citep*{loo82,mil04,bek07,sch06,sch08}. The second mechanism suggests that massive stellar clusters, such as Globular Clusters (GCs), spiral into the galactic centre and merge to form a dense nucleus \citep{tre75,cap93,cap08,ant12,ant13}. Observations show that both processes occur in nature and contribute to the formation of NSCs. In the latter scenario, GCs are expected to be totally disrupted by BH tidal forces, but some stars may be accelerated to high velocities and ejected in jets from the inner galactic regions \citep*{cap15,acs15}.

High velocity stars have been observed in the Galactic halo. Such high velocities may be gained thanks to several physical mechanisms, as three-body interactions among binary systems or with the massive black hole in the Galactic centre, or kicks due to supernovae explosions. High velocity stars can be divided in two different categories, i.e. runaway stars (RS) and hypervelocity stars (HVSs).

RSs, historically defined in the context of O and B stars \citep{hum47}, are Galactic halo stars with peculiar motions higher than $40$ km s$^{-1}$. Such young massive stars are not expected to be present far from star-forming regions and are thought to have travelled far from their birthplace. Two mechanisms are expected to produce RSs: supernova ejections and dynamical ejections \citep{sil11}. In the first scenario \citep{bla61,por00,sck06,prz08}, a RS has origin in a binary system, where it receives a velocity kick when its companion explodes as a supernova. In the dynamical ejection scenario \citep*{pov67}, stars are accelerated thanks to a three- or four-body interaction \citep*{leo90,gvr09,gva09,gva11,per12}. Observations show that both the ejection mechanisms operate in nature \citep{hoo01}, but, in any case, RSs velocities are under the Galaxy escape velocity.

HVSs are stars escaping the host Galaxy. \citet{hil88} was the first to predict theoretically their existence, while \citet{brw05} serendipitously discovered the first HVS in the outer halo. Hills' mechanism involves the tidal breakup of a binary passing close to a massive BH and has been investigated in literature by several authors \citep*{yut03,gua05,brm06,sar09,kob12,ros14}. Moreover, Hills' scenario predicts the existence of a population of stars orbiting in the inner Galactic regions around the central BH \citep*{gou03,gin06,per07}. Also other mechanisms have been proposed to explain the production of HVSs \citep{tuf09,brw15}, as the interaction of a SMBHB with a single star \citep*{gua05, bau06, ses06}, the arrival from another nearby galaxy \citep*{gua07,bon08,she08,per09,brw10} and supernova explosion in a close binary \citep*{zub13}.

Since high velocity stars production mechanisms involve different astrophysical phenomena, it would be possible to infer information about different branches of physics, as the physics of regions near massive BHs \citep*{gou03,ses07,oll08} and of supernovae \citep{por00,zub13}. Moreover, the study of the proper motions of such fast moving stars can improve the knowledge of the Galaxy gravitational potential shape and of its Dark Matter component \citep{gne05,yum07}.

Observations of high velocity and hypervelocity objects have usually been limited to high-mass, early-type stars due to obvious observational bias \citep*{brw09,brw10,brw14}. Nowadays, observers have started investigating low-mass high velocity stars \citep*{pal14,zho14,li15,vic15}, some of which are low-mass HVSs candidates. Moreover, the European ESA satellite \textit{Gaia} (http://www.cosmos.esa.int/web/gaia), along with \textit{Gaia-ESO} (http://www.gaia-eso.eu/), is expected to measure proper motions with an unprecedented precision, providing for a larger and less biased sample, and to find $\sim 100$ HVSs in a sample of $\sim 10^9$ stars.

The aim of this paper is to investigate the production mechanism of high velocity stars, which involves a GC and a SMBHB. Actually, when the orbit of a GC delivers it close to a BHB in the centre of its host galaxy due to dynamical friction, some stars are stripped from the cluster and may be ejected with high velocities. For these test cases, we assumed a total mass $M=10^8$ M$_\odot$ with the scope of better underlying the physical mechanism and of making comparison with our previous results in the case of a single BH \citep{cap15}.

The paper is organised as follows. In Section \ref{sec:meth} we outline and describe our approach to the study of the consequences of the GC-BHB interaction. In Section \ref{sec:res} the results of our scattering experiments are presented and discussed. Finally, in Section \ref{sec:con} we draw the conclusions.

\section{Method}
\label{sec:meth}
Our scattering experiments refer to the interaction of three different components: a SMBHB, a GC and a star. In our simulations the SMBHB centre of mass sits initially in the origin of the reference frame, while the GC follows an elliptical orbit at a relatively close distance around it. The assumption of close distance to the BH is motivated by the fact that the globular cluster is supposedly orbitally decayed by dynamical friction braking suffered in the inner dense galactic regions.

The SMBHB initial circular orbit has radius $a_h$. Therefore, for the heavier black hole the initial conditions are given by

\begin{equation}
r_{1,c}=\frac{m_2}{M}a_h\ \ \ ;\ \ \ v_{1,c}=\sqrt{\frac{G}{M a_h}}m_2,
\end{equation}

while for the lighter black hole by

\begin{equation}
r_{2,c}=\frac{m_1}{M}a_h\ \ \ ;\ \ \ v_{2,c}=\sqrt{\frac{G}{M a_h}}m_1,
\end{equation}

where $r_1$ and $r_2$ are the radii of the circular orbits of $m_1$ and $m_2$, respectively, around their centre of mass.

The mechanical energy (per unit mass) of the GC on a circular orbit of radius $r_c$, around the BHB center of mass (neglecting the stellar background), is

\begin{equation}\label{eqn:ener}
E_{c}\equiv \frac{1}{2}v_c^2-\frac{GM}{r_c}=-\frac{1}{2}\frac{GM}{r_c},
\end{equation}

where $v_c=(GM/r_c)^{1/2}$ is the circular velocity. Thereafter, taking into account that the angular momentum (per unit mass) of the GC for the circular orbit is $L_c=\sqrt{GM r_c}$, the pericenter ($r_-$) and apocenter ($r_+$) distances of the GC on orbits of same energy ($E_c$), but different angular momentum $0\leq L\leq L_c$, are given by 

\begin{equation}\label{periapo}
r_{\pm} = r_c\left(1\pm \sqrt{1-\left(\frac{L}{L_c}\right)^2}\right),
\end{equation}

where the $-$ sign gives the pericenter and the $+$ sign gives the apocenter. Therefore, by varying the ratio $\alpha=(L/L_c)^2$, we can compare the circular orbit with a set of orbits at same energy, but different eccentricity 

\begin{equation}\label{ecc}
e=\frac{r_+-r_-}{r_-+r_+} = \sqrt{1-\alpha}.
\end{equation}

Furthermore, in our simulations, the GC is assumed to have a \citet{plu11} mass profile

\begin{equation} \label{eqn:plumm}
M(r)=M_{GC}\frac{r^3}{{(r^2+b^2)}^{3/2}},
\end{equation}

where $M_{GC}$ is the total mass of the GC and $b$ its core radius. The test star is on circular orbit inside the GC sphere of influence.
 
The cartesian reference frame has been chosen as that with the $x$-axis along the line connecting the GC with the SMBHB center of mass and $y$-axis orthogonal, so that the $(x,y)$ frame is equiverse to the GC orbital revolution.

To summarize, the values of the relevant initial parameters have been set as follows (see also Table 1): 

\begin{itemize}
\item the total mass of the SMBHB is $M=10^8$ M$_{\odot}$;
\item the binary ratio $\nu=m_2/M$ assumes the values of $1/2$, $1/3$, $1/4$, $1/5$, $1/10$ and $1/20$;
\item the radius $r_{BHB}$ of the SMBHB initial circular orbit is set equal to $a_h$;
\item the initial phase $\Psi$ of the BHB is randomly generated;
\item the GC mass, $M_{GC}$, is fixed to $10^6$ M$_{\odot}$;
\item the GC core radius $b$ is set to $0.2$ pc;
\item the radius of the GC reference circular orbit is $r_0=10$ pc; 
\item the GC orbital eccentricity ranges from $e=0.71$ ($\alpha = 0.5$) to $e=0.95$ ($\alpha = 0.1$) and is parametrized varying $0.1\leq \alpha \leq 0.5$ at steps of $0.1$;
\item the GC orbits are coplanar with the SMBHB one; 
\item the test star mass, $m_{*}$, is set equal to $1$ M$_{\odot}$;
\item the test star circular orbit radii are equal to $r_L/4$, $r_L/6$, $r_L/8$, $r_L/10$, where $r_L=r_0(M_{GC}/3M)^{1/3}$ is the radius of the Hill's sphere;
\item the star initial position on the circular orbits is randomly generated;
\item the angles $\theta$, $\phi$, $\psi$, which determine the orientation of the star circular orbit in the GC reference frame, are randomly generated.
\end{itemize}

\begin{table}
\caption{The values of the binary mass ratio $\nu$ and of the binary circular orbit $r_{BHB}$.}
\centering
\begin{tabular}{c|c|}
\hline
$\nu$ & $r_{BHB}$ (pc) \\
\hline
1/20 & 0.16 \\
\hline
1/10 & 0.31 \\
\hline
1/5 & 0.62 \\
\hline
1/4 & 0.78 \\
\hline
1/3 & 1.04 \\
\hline
1/2 & 1.56 \\
\hline
\end{tabular}
\label{tab1}
\end{table}

The choice of the range of $\alpha$, and consequently of $e$, toward large values of $e$, is justified by the fact that the efficiency of the energy transfer on the test star orbiting the GC tends to vanish at eccentricities less than $\sim 0.6$.

Given the above set of initial parameters, we integrated the system of the differential equations of the 4-bodies (SMBHB, GC and star) motion

\begin{equation} \label{eqn:mot}
{\ddot{\textbf{r}}}_i=-G\sum\limits_{j\ne i}\frac{m_j(\textbf{r}_i-\textbf{r}_j)}{\left|\textbf{r}_i-\textbf{r}_j\right|^3},
\end{equation}

for $i=1, 2, 3, 4$, using the fully regularized algorithm of \citet{mik01}. The enormous range of variation of the involved masses requires a regularized algorithm, since any not-regularized direct summation code would fail in dealing with the close star-SMBHB interaction and would carry to a large energy error during the close encounter. Thanks to a transformed leapfrog method, combined with the Bulirsch-Stoer extrapolation method, the Mikkola's ARW code overcomes this problem and leads to extremely accurate integrations of the bodies trajectories \citep{bul66,mik99a,mik99b,mik06,mik08,hel10}. Thanks to the regularized algorithm, the fractional energy error is kept  below $10^{-10}$ over the whole integration time.

\section{Results}
\label{sec:res}

In our scattering experiments the test star orbiting the GC has three possible fates after the interaction with the SMBHB. Actually, the star can either remain bound to the GC, but on an orbit significantly perturbed respect to the original one, or can be captured by the SMBHB and starts orbiting around the galactic centre on precessing loops, or can be lost by the GC-SMBHB system.

The distinction among these three different situations is made by computing the mechanical energy of the star respect to the SMBHB and the GC after the scattering. If its energy respect to the GC remains negative, the star remains bound to the GC, while if this energy becomes positive, while the star energy respect to the BHB is negative, the star becomes bound to the binary. Finally, if both these energies are positive, the star is able to leave the SMBHB-GC system.

In this paper, we focus our attention on the stars ejected at high velocities after the close interaction with a SMBHB, studying the effects of the mass ratio and of the SMBHB orbital eccentricity.

\subsection{The role of the mass ratio}

\begin{figure}\label{fig:vel_dist}
\centering
\subfigure{\includegraphics[scale=0.7]{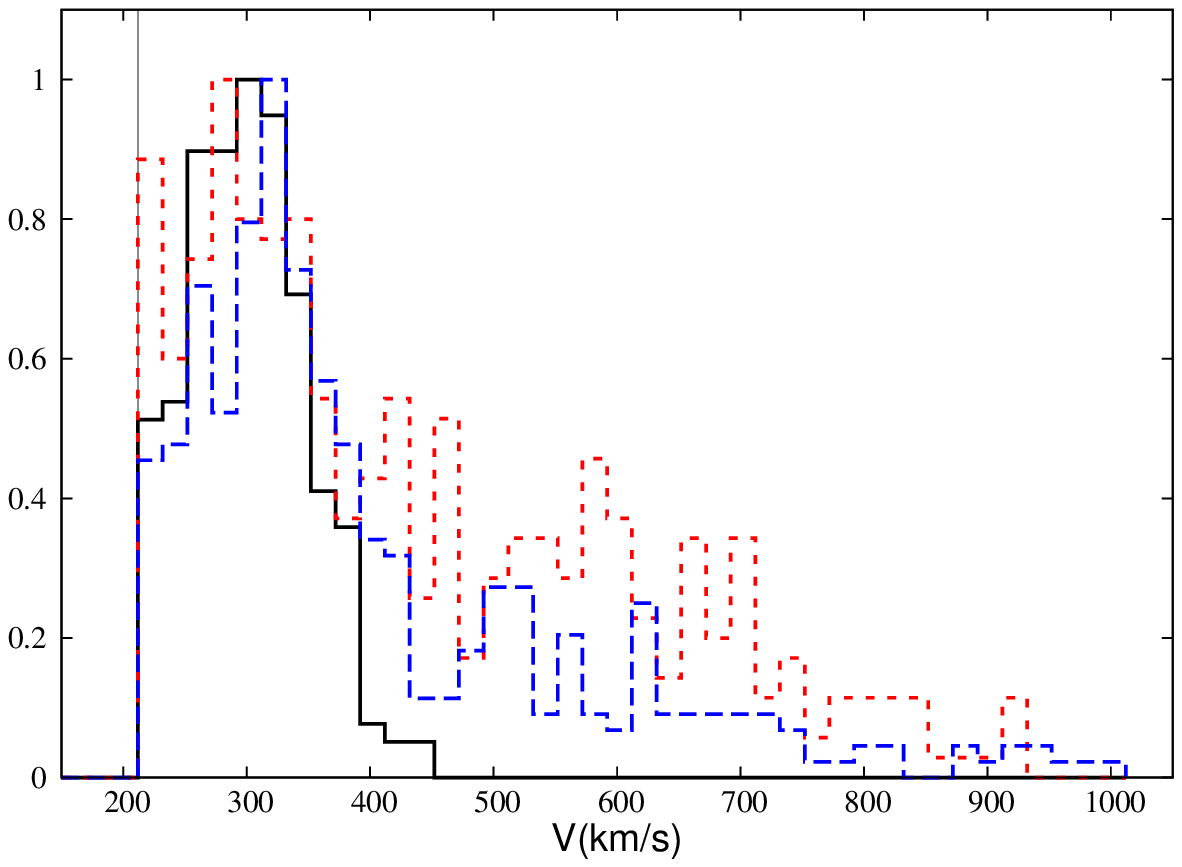}}
\subfigure{\includegraphics[scale=0.7]{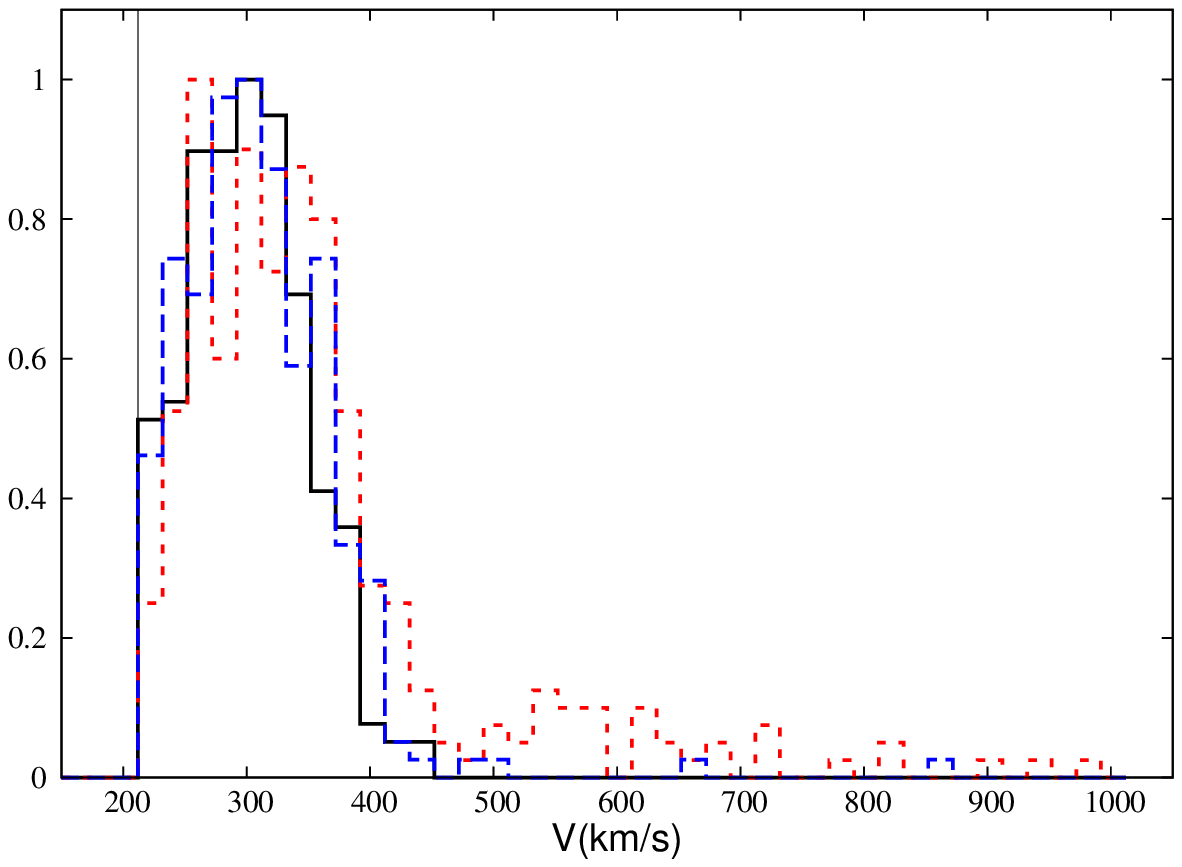}}
\caption{Comparison between the velocity distributions of escaping stars for different binary mass ratios and single Black Hole ($\nu=0$, black line). The top panel respresents the distributions for high mass ratios ($\nu=1/3$, red line, and $\nu=1/5$, blue line), while the bottom panel low mass ratios ($\nu=1/10$, red line, and $\nu=1/20$, blue line). The distributions are cut on the left side at $212$ km s$^{-1}$, which corresponds to the escape velocity respect to the system at $20$ pc.}
\end{figure}

From our scattering experiments, it is possible to derive the velocity profile of the ejected stars. In the case of single massive BH, the velocity distribution is narrow and peaked at small velocities, depending on the GC core radius. Actually, when the GC has a Plummer profile (Eq. \ref{eqn:plumm}), for the set of parameters chosen in this work, the initial gravitational energy of the star is $\sim GM_{GC}/b$ (in our simulations $b=0.2$ pc). Therefore, the distribution is peaked at a lower velocity respect to the case of point-mass GC \citep{cap15}. The ejection velocity roughly depends on the initial amount of star gravitational energy $E_{*,g}$ and on the pericentre $r_p$ of the GC orbit. Moreover, the dispersion of the nearly Gaussian distribution is determined, besides by the different (randomly generated) angles and initial conditions, by the relative inclination $\gamma$ between the angular momentum of the cluster and that of the test star. Actually, if $\gamma \sim 0$ during the close interaction, i.e. near the pericentre of the GC orbit respect to the BHB centre of mass, the ejection velocity of the star will be higher than the cases in which $\gamma \gg 0$. Fig. 1 shows the resulting nearly Gaussian velocity distribution, for all the values of $\alpha$, peaked at $\sim 300$ km s$^{-1}$ in the case of single BH (black line). 

The situation is different when a scattering with a BHB is taken into account. Actually, the ejection velocity depends not only on $E_{*,g}$, $r_p$ and $\gamma$, but also on $\nu$, and so $a_h$. The generic star of an infalling GC is able to exchange energy with the binary through the gravitational slingshot, which enhances its ejection velocity. On the other hand, the energy transfer makes the SMBHB shrink and reduces the apocentre of its orbit. Fig. 1 shows the resulting velocity distribution for BHB for different values of $\nu$. If the mass ratio is high (top panel), a considerable tail in the distribution up to $\sim 1000$ km s$^{-1}$ is produced. On the contrary, if the mass ratio is low (bottom panel), the differences between the distributions are not so pronounced. In particular, if $\nu\lesssim 1/20$, the velocity distributions for single and binary BH are essentially similar.

Velocity distributions show that for high values of $\nu$, there is a considerable fraction of the ejected stars, which acquire significant high velocities. Actually. for $\nu \gtrsim 1/5$, the huge tail of the distribution extends up to $\sim 1000$ km s$^{-1}$. Therefore, a considerable fraction of ejected stars is unbound not only respect to the SMBHB-GC system, but also respect to the host galaxy, becoming HVSs. The escape velocity (at $20$ pc as justified in \citet{cap15}) for an elliptical galaxy, as NGC 3377 \citep{mar03}, of total mass $M_{E}=7.81 \times 10^{10}$ M$_{\odot}$, is $418$ km s$^{-1}$ \citep{her90}, while for a spiral galaxy, of total mass $M_{S}=6.60 \times 10^{11}$ M$_{\odot}$ \citep*{wan92,rwn94,krn08,lin14}, is $759$ km s$^{-1}$ \citep*{miy75,bin81,her90,fuj09}. Fig. 1 shows that a not-negligible fraction of ejected stars is beyond the local escape velocity. The fate of these stars is to escape the SMBHB-GC system, to travel across the halo and to leave the host galaxy.

The introduction of a secondary BH, comparable in mass with the primary, leads to a peculiar distribution velocity for the stars ejected at high velocities. In principle, these distributions can be used, when data for proper motions and radial velocities will be available for galaxies whose central object(s) total mass is $M=10^8$ M$_{\odot}$, to distinguish if the central object is a single BH or a BHB.

The effect of the presence of a secondary BH is not limited to the velocity distribution of ejected stars. Actually, also the branching ratios of ejected stars, i.e. the probabilities that the system BHB-GC loses stars, depend on the mass ratio $\nu$. Fig. 2 shows the branching ratios for the case of single and binary BH. In the case of single BH ($\nu=0$), the branching ratio is $0.267$ for the set of parameter studied in this work. For low values of the mass ratio, i.e. $\nu \lesssim 1/20$, the branching ratio remains nearly constant, while for higher values, it is an increasing function of $\nu$. Therefore, only if the mass ratio is sufficiently high, the probability increases respect to the case of single BH.

In conclusion, the effect of the binariety, of course if $\nu \gtrsim 1/20$, is dual. On the one hand it enhances the probability of stars ejections, on the other it produces a considerable tail in the velocity distribution, leading to the production of HVSs.

\begin{figure} \label{fig:bratio}
\centering
\includegraphics[scale=0.7]{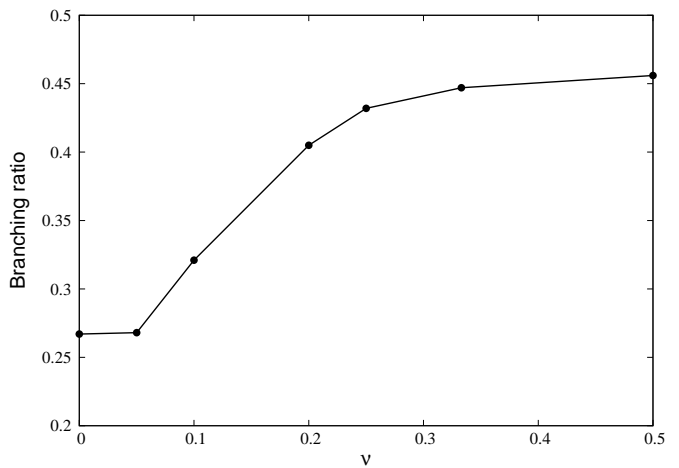}
\caption{Branching ratios for high velocity stars as function of the binary mass ratio $\nu$.}
\end{figure}

\subsection{The role of the binary eccentricity}

In order to explore the role of the BHB orbital eccentricity $\eta$, we performed the same set of simulations, in the case of mass ratio $\nu=1/4$, for $\eta=0.25$, $0.50$, $0.75$ elliptical orbits. The angular momenta are chosen such that, as in the case of GC, the energy of elliptical orbits is equal to the energy of the reference circular orbit ($\eta=0$) of radius $a_h=0.78$ pc. In this case, the initial conditions, assuming that the black holes start their orbital motion at the apocentre, are

\begin{equation}
r_1=(1+\eta)r_{1,c}\ \ \ ;\ \ \ v_1=F(\eta)v_{1,c}=\sqrt{\frac{1-\eta}{1+\eta}} v_{1,c},
\end{equation}

for the heavier black hole $m_1$, while for the lighter black hole $m_2$ 

\begin{equation}
r_2=(1+\eta)r_{2,c}\ \ \ ;\ \ \ v_2=F(\eta)v_{2,c}=\sqrt{\frac{1-\eta}{1+\eta}} v_{2,c}.
\end{equation}

On the contrary, if the black holes start their orbital motion at the pericentre, $r_{1-2}=(1-\eta)r_{1-2,c}$ and $v_{1-2}=F^{-1}(\eta)v_{1-2,c}$. Finally, for this set of simulations, the angle $\zeta$ between the semi-major axis of the SMBHB and the x-axis of the cartesian reference frame is randomly generated.

Fig. 3 shows the resulting velocity distributions for a BHB with mass ratio $\nu=1/4$ and different orbital eccentricities. It is clear that the tail in the velocity distribution is produced independently on $\eta$, with velocities up to $\sim 1000$ km s$^{-1}$. However, different eccentricities do not change the shape of the distribution function respect to the case of the circular orbit. Therefore, these results suggest that, although the different shape of the binary orbit, the energy exchange between the generic star of the GC and the BHB is independent on the eccentricity, but depends only on the initial total energy of the binary.

For what regards the branching ratios, Tab. 2 shows the resulting branching ratios of our scattering experiments as function of $\eta$. From Tab. 2, it is clear that the branching ratio remains nearly constant for different eccentricities. Therefore, not only the shape of the velocity distribution is preserved, but also the probability of stars ejection.

In conclusion, the BHB orbital eccentricity does not effect neither the shape of the velocity distribution nor the branching ratio of ejected stars. Therefore, our simulations suggest that the features of the ejected stars depend only on the binary mass ratio $\nu$, but not on the binary eccentricity.

\begin{figure} \label{fig:vel_ecc}
\centering
\includegraphics[scale=0.7]{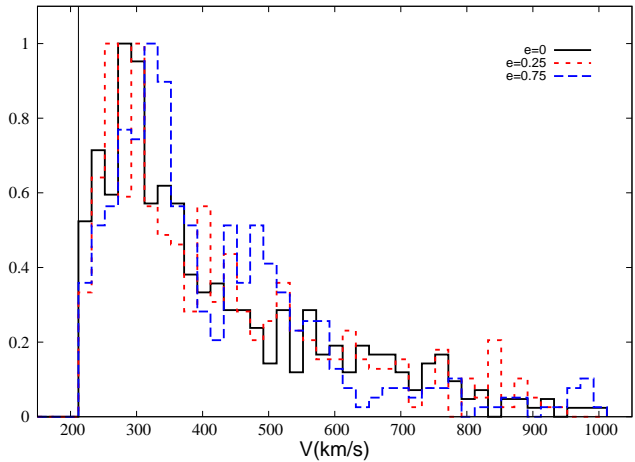}
\caption{Comparison between the velocity distributions of escaping stars for a SMBHB with mass ratios $\nu=1/4$ and different eccentricities ($\eta=0$, $0.25$, $0.75$). The distributions are cut on the left side at $212$ km s$^{-1}$, which corresponds to the escape velocity respect to the system at $20$ pc.}
\end{figure}

\begin{table}
\caption{The values of the Branching Ratios (BR) for $\nu=1/4$ and different BHB orbital eccentricities $\eta$.}
\centering
\begin{tabular}{c|c|}
\hline
$\eta$ & BR \\
\hline
0 & 0.432 \\
\hline
0.25 & 0.403 \\
\hline
0.50 & 0.440 \\
\hline
0.75 & 0.425 \\
\hline
\end{tabular}
\label{tab1}
\end{table}

\section{Conclusions}
\label{sec:con}

The existence of high velocity stars \citep{sil11} has been explained thanks to a dynamical ejection mechanism \citep{pov67,gva09} or a supernova ejection mechanism \citep{bla61,por00}, which are able to accelerated stars up to several hundreds km s$^{-1}$. On the other hand, HVSs \citep{hil88} require the presence of a massive BH, or massive BHB, due to their extreme velocities up to thousands km s$^{-1}$ \citep{yut03,brw15}.

In this paper, we extended the recently study \citet{cap15}, referred to the case of a single BH. Here we generalised the study to the ejection of high velocity stars due to the close passage of a massive GC near a massive BHB. In the frame of the $\Lambda$CDM cosmological model, SMBHBs are a natural consequence of the hierarchical paradigm \citep*{beg80,vol03}. Actually, galaxies may experience multiple mergers during their lifetime and, if more than one of them host a massive BH, the formation of a Black Hole Binary is a natural result.

In our study, we assumed a total mass $M=10^8$ M$_{\odot}$ of the SMBHB to have a direct comparison with previous scattering experiments with a single BH \citep{cap15}. The underlying mechanism is a four-body interaction, where the bodies are the SMBHB, the GC (of mass $10^6$ M$_{\odot}$ in our study) and a generic test star ($1$ M$_{\odot}$) belonging to the cluster. We performed a series of high precision scattering experiments in order to investigate the ejection probability, after a close interaction with a SMBHB, and the distribution velocity of the ejected stars.

In \citet{cap15} results, the velocity distribution was nearly Gaussian, narrow and peaked at low values in dependence on the GC core radius. The ejection velocity is roughly determined by the initial amount of star gravitational energy $E_{*,g}$ and by the pericentre $r_p$ of the GC orbit, while the dispersion of the velocity distribution by the different initial conditions and by the relative inclination angle between the angular momentum of the cluster and that of the test star. Moreover, when the interaction occurs with a SMBHB, GC stars are able to exchange energy through gravitational slingshots, which enhances its ejection velocity and produces a tail in the distribution, which depends on the mass ratio $\nu$ of the BHB. Actually, the higher is the mass ratio, the more extended the distribution toward high velocities will be. At the same time, also the branching ratio for ejected stars depends on the mass ratio $\nu$. While for low values of $\nu$, the branching ratio remains nearly at the value of the single BH case ($0.267$), for $\nu \gtrsim 1/20$, it is an increasing function of the mass ratio.

Finally, we performed the same set of simulations in the case of mass ratio $\nu=1/4$, but assuming different eccentricities $\eta$. The shape of the velocity distribution and the values of the branching ratios show that eccentricity has not a substantial effect on the results. Therefore, our simulations suggest that the features of the ejected stars depend only on the binary mass ratio.

In conclusion, the effect of the binariety (if $\nu \gtrsim 1/20$) is dual, since it enhances the probability of stars ejections and produces an extended tail in the velocity distribution, corresponding to the production of HVSs. Finally, we suggest that the infall of a GC on a SMBHB could enhance the energy loss by the BHB and conduce them to the final stage, where gravitational waves emission is the main mechanism to cause the energy loss of the binary.

\section*{Acknowledgments}

We thank S. Mikkola for making available to us his ARW code and for useful discussions about his use.


\begin{thebibliography}{99}
\bibitem[\protect\citeauthoryear{Antonini}{2013}]{ant13} Antonini F., 2013, ApJ, 763.1, 62
\bibitem[\protect\citeauthoryear{Antonini et al.}{2012}]{ant12} Antonini F., Capuzzo-Dolcetta R., Mastrobuono-Battisti A., Merritt D., 2012, ApJ, 750, 111
\bibitem[\protect\citeauthoryear{Arca-Sedda, Capuzzo-Dolcetta, Spera}{Arca-Sedda et al.}{2015}]{acs15} Arca-Sedda M., Capuzzo-Dolcetta R., Spera M., 2015, preprint (arXiv:1510.01137v1)
\bibitem[\protect\citeauthoryear{Baumgardt, Gualandris \& Portegies Zwart}{Baumgardt et al.}{2006}]{bau06} Baumgardt H., Gualandris A., Portegies Zwart S.F., 2006, J. Phys.: Conf. Ser., 54.1, 301
\bibitem[\protect\citeauthoryear{Begelman, Blandford \& Rees}{Begelman et al.}{1980}]{beg80} Begelman M.C., Blandford R.D., Rees M.J., 1980, Nature, 287, 307
\bibitem[\protect\citeauthoryear{Bekki}{2007}]{bek07} Bekki K. 2007, Publ. Astron. Soc. Austr., 24, 77B
\bibitem[\protect\citeauthoryear{Berczik et al.}{2006}]{ber06} Berczik P., Merrit D., Spurzem R., Bischof H.-P., 2006, ApJ Lett., 642.1, L21
\bibitem[\protect\citeauthoryear{Binney}{1981}]{bin81} Binney J., 1981, MNRAS, 196, 455
\bibitem[\protect\citeauthoryear{Blaauw}{1961}]{bla61} Blaauw A., 1961, Bull. Astron. Inst. Netherlands, 15, 265
\bibitem[\protect\citeauthoryear{Bonanos et al.}{2008}]{bon08} Bonanos A.Z., Lopez-Morales M., Hunter I., Ryans R.S.I., 2008, ApJ, 675, L77
\bibitem[\protect\citeauthoryear{Bromley et al.}{2006}]{brm06} Bromley B.C., Kenyon S.J., Geller M.J., Barcikowski E., Brown W.R., Kurtz M.J., 2006, ApJ, 653, 1194
\bibitem[\protect\citeauthoryear{Brown}{2015}]{brw15} Brown W.R., 2015, Annu. Rev. Astron. Astrophys., 53, 15
\bibitem[\protect\citeauthoryear{Brown, Geller \& Kenyon}{Brown et al.}{2009}]{brw09} Brown W.R., Geller M.J., Kenyon S.J., 2009, ApJ, 690.2, 1639
\bibitem[\protect\citeauthoryear{Brown, Geller \& Kenyon}{Brown et al.}{2014}]{brw14} Brown W.R., Geller M.J., Kenyon S.J., 2014, ApJ, 787, 89
\bibitem[\protect\citeauthoryear{Brown et al.}{2005}]{brw05} Brown W.R., Geller M.J., Kenyon S.J., Kurtz M.J., 2005, ApJ Lett., 622, L33
\bibitem[\protect\citeauthoryear{Brown et al.}{2010}]{brw10} Brown W.R. et al., 2010, ApJ Lett., 719, L23
\bibitem[\protect\citeauthoryear{Bulirsch \& Stoer}{1966}]{bul66} Bulirsch R., Stoer J., 1966, Numer. Math., 8, 1
\bibitem[\protect\citeauthoryear{Capuzzo-Dolcetta}{1993}]{cap93} Capuzzo-Dolcetta R., 1993, ApJ, 415, 616
\bibitem[\protect\citeauthoryear{Capuzzo-Dolcetta \& Fragione}{2015}]{cap15} Capuzzo-Dolcetta R., Fragione G., 2015, MNRAS, 2015, 454.3, 2677-2690
\bibitem[\protect\citeauthoryear{Capuzzo-Dolcetta \& Miocchi}{2008}]{cap08} Capuzzo-Dolcetta R., Miocchi P., MNRAS Lett., 388, L69
\bibitem[\protect\citeauthoryear{Carollo et al.}{1997}]{car97} Carollo C. M., Stiavelli M., de Zeeuw P.T., Mack J., 1997, AJ, 114, 2366
\bibitem[\protect\citeauthoryear{Carollo, Stiavelli \& Mack}{Carollo et al.}{1998}]{car98} Carollo C.M., Stiavelli M., Mack J. 1998, AJ, 116, 68
\bibitem[\protect\citeauthoryear{Chandrasekhar}{1943}]{cha43} Chandrasekhar S., AJ, 1943, 97, 255
\bibitem[\protect\citeauthoryear{Chen et al.}{2009}]{che09} Chen X., Madau P., Sesana A., Liu F.K., 2009, ApJ, 697, L149
\bibitem[\protect\citeauthoryear{Chen et al.}{2011}]{che11} Chen X., Sesana A., Madau P., Liu F.K., 2011, ApJ, 729, 13
\bibitem[\protect\citeauthoryear{C{\^o}t\'e et al.}{2006}]{cot06} C{\^o}t\'e P. et al., 2006, ApJ Suppl. Ser., 165, 57
\bibitem[\protect\citeauthoryear{Deane et al.}{2014}]{dea14} Deane R.P. et al., 2014, Nature, 511, 57
\bibitem[\protect\citeauthoryear{Dubois Volonteri \& Silk}{Dubois et al.}{2014}]{dub14} Dubois Y., Volonteri M., Silk J., 2014, MNRAS, 440, 1590
\bibitem[\protect\citeauthoryear{Fabbiano et al.}{2011}]{fab11} Fabbiano G., Wang J., Elvis M., Risaliti G., 2011, Nature, 477, 431
\bibitem[\protect\citeauthoryear{Fujita}{2009}]{fuj09} Fujita Y., 2009, ApJ, 691, 1050
\bibitem[\protect\citeauthoryear{Ginsburg \& Loeb}{2006}]{gin06} Ginsburg I., Loeb A., 2006, MNRAS, 368, 221
\bibitem[\protect\citeauthoryear{Gnedin et al.}{2005}]{gne05} Gnedin O.Y., Gould A., Miralda-Escudé J., Zentner A.R., 2005, ApJ, 634, 344.
\bibitem[\protect\citeauthoryear{Goicovic et al.}{2015}]{goi15} Goicovic F.G. et al., 2015, preprint (arXiv:1507.05596)
\bibitem[\protect\citeauthoryear{Gould \& Quillen}{2003}]{gou03} Gould A., Quillen A.C., 2003, ApJ, 592, 935
\bibitem[\protect\citeauthoryear{Gualandris \& Portegies Zwart}{2007}]{gua07} Gualandris A., Portegies Zwart S.F., 2007, MNRAS Lett., 376.1, L29
\bibitem[\protect\citeauthoryear{Gualandris, Portegies Zwart \& Sipior}{Gualandris et al.}{2005}]{gua05} Gualandris A., Portegies Zwart S.F., Sipior M.S., 2005, MNRAS, 363, 223
\bibitem[\protect\citeauthoryear{Gvaramadze}{2009}]{gvr09} Gvaramadze V.V., 2009, MNRAS, 395, L85
\bibitem[\protect\citeauthoryear{Gvaramadze \& Gualandris}{2011}]{gva11} Gvaramadze V.V., Gualandris A., 2011, MNRAS, 410, 304
\bibitem[\protect\citeauthoryear{Gvaramadze, Gualandris \& Portegies Zwart}{Gvaramadze et al.}{2009}]{gva09} Gvaramadze V.V., Gualandris A., Portegies Zwart S.F., 2009, MNRAS, 396, 570
\bibitem[\protect\citeauthoryear{Hellstr\"{o}m \& Mikkola}{2010}]{hel10} Hellstr\"{o}m C., Mikkola S., 2010, Celest. Mech. Dyn. Astron., 106, 143
\bibitem[\protect\citeauthoryear{Hernquist}{1990}]{her90} Hernquist L., 1990, ApJ, 356, 359
\bibitem[\protect\citeauthoryear{Hills}{1988}]{hil88} Hills J.G., 1988, Nature, 331, 687
\bibitem[\protect\citeauthoryear{Hoogerwerf, de Bruijne \& de Zeeuw}{Hoogerwerf et al.}{2001}]{hoo01} Hoogerwerf R., de Bruijne J.H.J., de Zeeuw P.T., 2001, A \& A, 365, 49
\bibitem[\protect\citeauthoryear{Humason \& Zwicky}{1947}]{hum47} Humason M.L., Zwicky F., 1947, ApJ, 105, 85
\bibitem[\protect\citeauthoryear{Khan et al.}{2013}]{kha13} Khan F.M., Holley-Bockelmann K., Berczik P., Just A., 2013, ApJ, 773.2, 100
\bibitem[\protect\citeauthoryear{Kobayashi et al.}{2012}]{kob12} Kobayashi S., Hainick Y., Sari R., Rossi E.M., 2012, ApJ, 670, 747
\bibitem[\protect\citeauthoryear{Komossa et al.}{2003}]{kom03} Komossa S. et al., 2003, ApJ, 582, L15
\bibitem[\protect\citeauthoryear{Kormendy \& Ho}{2013}]{kor13} Kormendy J., Ho L.C., 2013, Annu. Rev. Astron. Astrophys., 51, 511
\bibitem[\protect\citeauthoryear{Kornreich \& Lovelace}{2008}]{krn08} Kornreich D.A., Lovelace R.V.E., 2008, ApJ, 681, 104
\bibitem[\protect\citeauthoryear{Leonard \& Duncan}{1990}]{leo90} Leonard P.J.T., Duncan M.J., 1990, AJ, 99, 608
\bibitem[\protect\citeauthoryear{Li et al.}{2015}]{li15} Li Y. et al., 2015, preprint (arXiv:1506.01818v2)
\bibitem[\protect\citeauthoryear{Lingam}{2014}]{lin14} Lingam M., 2014, Astrophys. Space Sci., 354, 561
\bibitem[\protect\citeauthoryear{Liu \& Chen}{2013}]{liu13} Liu F.K., Chen X., 2013, ApJ, 767, 18
\bibitem[\protect\citeauthoryear{Liu, Li \& Chen}{Liu et al.}{2009}]{liu09} Liu F.K., Li S., Chen X., 2009, ApJ, 706, L133
\bibitem[\protect\citeauthoryear{Liu, Li \& Komossa}{Liu et al.}{2014}]{liu14} Liu F.K., Li S., Komossa K., 2014, ApJ, 786, 103
\bibitem[\protect\citeauthoryear{Loose, Kruegel \& Tutukov}{Loose et al.}{1982}]{loo82} Loose H.H., Kruegel E., Tutukov A., 1982, A\& A, 105, 342
\bibitem[\protect\citeauthoryear{Marconi \& Hunt}{2003}]{mar03} Marconi A., Hunt L.K., 2003, ApJ Lett. 589, L21
\bibitem[\protect\citeauthoryear{Matthews \& Gallagher}{1997}]{mat97} Matthews L.D., Gallagher J.S., 1997, AJ, 114, 1899
\bibitem[\protect\citeauthoryear{Mayer at al.}{2007}]{may07} Mayer L. at al., 2007, Science, 316, 1874
\bibitem[\protect\citeauthoryear{Merritt, Mikkola \& Szell}{Merrit et al.}{2007}]{mer07} Merritt D., Mikkola S., Szell A., 2007, ApJ, 671, 53
\bibitem[\protect\citeauthoryear{Merritt \& Milosavljevi{\'c}}{2005}]{mer05} Merritt D., Milosavljevi{\'c} M., 2005, Living Rev. Relativ., 8, 8
\bibitem[\protect\citeauthoryear{Mikkola \& Aarseth}{2001}]{mik01} Mikkola S., Aarseth S., 2001, Celest. Mech. Dyn. Astron., 84, 343
\bibitem[\protect\citeauthoryear{Mikkola \& Merritt}{2006}]{mik06} Mikkola S., Merritt D., 2006, MNRAS, 372, 219
\bibitem[\protect\citeauthoryear{Mikkola \& Merritt}{2008}]{mik08} Mikkola S., Merritt D., 2008, AJ, 135, 2398
\bibitem[\protect\citeauthoryear{Mikkola \& Tanikawa}{1999a}]{mik99a} Mikkola S., Tanikawa K., 1999a, Celest. Mech. Dyn. Astron., 74, 287
\bibitem[\protect\citeauthoryear{Mikkola \& Tanikawa}{1999b}]{mik99b} Mikkola S., Tanikawa K., 1999b, MNRAS, 310, 745
\bibitem[\protect\citeauthoryear{Milosavljevi{\'c}}{2004}]{mil04} Milosavljevi{\'c} M., 2004, ApJ Lett., 605, 13
Milosavljevi´
\bibitem[\protect\citeauthoryear{Miyamoto \& Nagai}{1975}]{miy75} Miyamoto M., Nagai R., 1975, Publ. Astron. Soc. Jpn., 27, 533
\bibitem[\protect\citeauthoryear{O'Leary \& Loeb}{2008}]{oll08} O'Leary R. M., Loeb A., 2008, MNRAS, 383, 86
\bibitem[\protect\citeauthoryear{Palladino et al.}{2014}]{pal14} Palladino L.E. et al., 2014, ApJ, 780, 7
\bibitem[\protect\citeauthoryear{Perets}{2009}]{per09} Perets H.B., 2009, ApJ, 698, 1330
\bibitem[\protect\citeauthoryear{Perets, Hopman \& Alexander}{Perets et al.}{2007}]{per07} Perets H.B., Hopman C., Alexander T., 2007, ApJ, 656, 709
\bibitem[\protect\citeauthoryear{Perets \& Subr}{2012}]{per12} Perets H.B., Subr L., 2012, ApJ, 751, 133
\bibitem[\protect\citeauthoryear{Peters}{1964}]{pet64} Peters P.C., 1964, Phys. Rev. B, 136, 1224
\bibitem[\protect\citeauthoryear{Plummer}{1911}]{plu11} Plummer H.C., 1911, MNRAS, 71, 140
\bibitem[\protect\citeauthoryear{Portegies Zwart}{2000}]{por00} Portegies Zwart S.F., 2000, ApJ, 544, 437
\bibitem[\protect\citeauthoryear{Poveda, Ruiz \& Allen}{Poveda et al.}{1967}]{pov67} Poveda A., Ruiz J., Allen C., 1967, Bol. Obser. Tonantzintla y Tacubaya, 4, 86
\bibitem[\protect\citeauthoryear{Przybilla et al.}{2008}]{prz08} Przybilla N., Nieva M.F., Heber U., Butler K., 2008, ApJ Lett., 684, L103
\bibitem[\protect\citeauthoryear{Quinlan}{1996}]{qui96} Quinlan G.D., 1996, New Astron., 1, 35
\bibitem[\protect\citeauthoryear{Rodriguez et al.}{2006}]{rod06} Rodriguez C. et al., 2006, ApJ, 646, 49
\bibitem[\protect\citeauthoryear{Rosado \& Sesana}{2014}]{rss14} Rosado P.A., Sesana A., 2014, MNRAS, 439, 3986
\bibitem[\protect\citeauthoryear{Ro{\v{s}}kar at al.}{2015}]{ros15} Ro{\v{s}}kar R. at al., 2015, MNRAS, 449, 494
\bibitem[\protect\citeauthoryear{Rossi, Kobayashi \& Sari}{Rossi et al.}{2014}]{ros14} Rossi E.M., Kobayashi S., Sari R., 2014, ApJ, 795.2, 125
\bibitem[\protect\citeauthoryear{Rownd, Dickey \& Helou}{Rownd et al.}{1994}]{rwn94} Rownd B.K., Dickey J.M., Helou G. 1994, AJ, 108, 1638
\bibitem[\protect\citeauthoryear{Sari, Kobayashi \& Rossi}{Sari et al.}{2009}]{sar09} Sari R., Kobayashi S., Rossi E.M., 2010, ApJ, 708, 605
\bibitem[\protect\citeauthoryear{Scheck et al.}{2006}]{sck06} Scheck L., Kifonidis K., Janka H.-T., M\"{u}ller E., 2006, A \& A, 457, 963
\bibitem[\protect\citeauthoryear{Schinnerer et al.}{2006}]{sch06} Schinnerer E. et al., 2006, ApJ, 649, 181
\bibitem[\protect\citeauthoryear{Schinnerer et al.}{2008}]{sch08} Schinnerer E. et al., 2008, ApJ Lett., 684, 21
\bibitem[\protect\citeauthoryear{Sesana}{2013}]{ses13} Sesana A., 2013, Class. Quant. Grav., 30.24, 244009
\bibitem[\protect\citeauthoryear{Sesana et al.}{2005}]{ses05} Sesana A. et al., 2005, ApJ, 623.1, 23
\bibitem[\protect\citeauthoryear{Sesana et al.}{2014}]{ses14} Sesana A., Barausse E., Dotti M., Rossi E.M., 2014, ApJ, 794.2, 104
\bibitem[\protect\citeauthoryear{Sesana, Haardt \& Madau}{Sesana et al.}{2006}]{ses06} Sesana A., Haardt F., Madau P., 2006, ApJ, 651.1, 392
\bibitem[\protect\citeauthoryear{Sesana, Haardt \& Madau}{Sesana et al.}{2007}]{ses07} Sesana A., Haardt F., Madau P., 2007, MNRAS Lett., 379, L45
\bibitem[\protect\citeauthoryear{Sherwin, Loeb \& O'Leary}{Sherwin et al.}{2008}]{she08} Sherwin B., Loeb A., O'Leary R., 2008, MNRAS, 386, 1179
\bibitem[\protect\citeauthoryear{Silva \& Napiwotzki}{2011}]{sil11} Silva M.D.V., Napiwotzki R., 2011, MNRAS, 411, 2596
\bibitem[\protect\citeauthoryear{Tremaine, Ostriker \& Spitzer}{Tremaine et al.}{1975}]{tre75} Tremaine S.D., Ostriker J.P., Spitzer L.J., 1975, ApJ, 196, 407
\bibitem[\protect\citeauthoryear{Turner et al.}{2012}]{tur12} Turner M.L. et al., 2012, ApJ Suppl. Ser., 203, 5
\bibitem[\protect\citeauthoryear{Tutukov \& Federova}{2009}]{tuf09} Tutukov A.V., Fedorova A.V., 2009, Astron. Rep., 53.9, 839
\bibitem[\protect\citeauthoryear{Vickers, Smith \& Grebel}{Vickers et al.}{2015}]{vic15} Vickers J.J., Smith M.C., Grebel E.K., 2015, AJ, 150.3, 77
\bibitem[\protect\citeauthoryear{Volonteri, Haardt \& Madau}{Volonteri et al.}{2003}]{vol03} Volonteri M., Haardt F., Madau P., 2003, ApJ, 582, 559
\bibitem[\protect\citeauthoryear{Wang, Sulkanen \& Lovelace}{Wang et al.}{1992}]{wan92} Wang J.C.L., Sulkanen M.E., Lovelace R.V.E., 1992, ApJ, 390, 46
\bibitem[\protect\citeauthoryear{Wegg \& Nate Bode}{2011}]{weg11} Wegg C. \& Nate Bode J., 2011, ApJ, 738, L8
\bibitem[\protect\citeauthoryear{Yu}{2002}]{yu02} Yu Q., 2002, MNRAS, 331, 935
\bibitem[\protect\citeauthoryear{Yu \& Madau}{2007}]{yum07} Yu Q., Madau P., 2007, MNRAS, 379, 1293
\bibitem[\protect\citeauthoryear{Yu \& Tremaine}{2003}]{yut03} Yu Q., Tremaine S., 2003, ApJ, 599, 1129
\bibitem[\protect\citeauthoryear{Zhong et al.}{2014}]{zho14} Zhong J. et al., 2014, ApJ, 789, L2
\bibitem[\protect\citeauthoryear{Zubovas, Wynn \& Gualandris}{2013}]{zub13} Zubovas K., Wynn G.A., Gualandris A., 2013, ApJ, 771.2, 118
\end{thebibliography}
\end{document}